# First-Principles Molecular Dynamics Investigation of the Atomic-Scale Energy Transport: From Heat Conduction to Thermal Radiation

Pengfei Ji and Yuwen Zhang[1]

Department of Mechanical and Aerospace Engineering

University of Missouri

Columbia, MO 65211, USA

## Abstract

First-principles molecular dynamics simulation based on a plane wave/pseudopotential implementation of density functional theory is adopted to investigate atomic scale energy transport for semiconductors (silicon and germanium). By imposing thermostats to keep constant temperatures of the nanoscale thin layers, initial thermal non-equilibrium between the neighboring layers is established under the vacuum condition. Models with variable gap distances with an interval of lattice constant increment of the simulated materials are set up and statistical comparisons of temperature evolution curves are made. Moreover, the equilibration time from non-equilibrium state to thermal equilibrium state of different silicon or/and germanium layers combinations are calculated. The results show significant distinctions of heat transfer under different materials and temperatures combinations. Further discussions on the equilibrium time are made to explain the simulation results. As the first work of the atomic scale energy transport spanning from heat conduction to thermal radiation, the simulation results here highlights the promising application of the first-principles molecular dynamics in thermal engineering.

---
[1] Corresponding author. Email: zhangyu@missouri.edu

# Nomenclature

| | |
|---|---|
| $\boldsymbol{a}$ | lattice vector, Bohr = 0.52917720859 Å |
| $\boldsymbol{b}$ | reciprocal lattice vector, $Bohr^{-1}$ |
| $c_0$ | speed of light in vacuum, m/s |
| $E$ | incident energy, J |
| $E_{ions}$ | interaction energy of the bare nuclear charges, Ha = 27.21138386 eV |
| $E^{KS}$ | Kohn-Sham energy, Ha |
| $E_{xc}$ | exchange functional, Ha |
| $f_i$ | integer occupation number |
| $f_G^{PW}$ | plane wave basis |
| $\boldsymbol{g}$ | a vector with triple of integer values, Bohr |
| $\boldsymbol{G}$ | reciprocal space vector, $Bohr^{-1}$ |
| $\boldsymbol{h}$ | lattice matrix |
| $\boldsymbol{k}$ | wave vector |
| $M$ | atomic mass, a.u. = 0.00054857990943 a.m.u |
| n | index of refraction |
| $\boldsymbol{R}$ | atomic position vector, Bohr |
| $\boldsymbol{R}^N$ | union of all the atomic position vectors |
| $s$ | coordinates scalar |
| $T_s$ | kinetic energy of a non-interacting system, Ha |
| $U$ | potential, Ha |
| $V_{ext}$ | external potential, Ha |
| $V_H$ | Hartree potential, Ha |

Greek

| | |
|---|---|
| $\varepsilon^{KS}$ | extended Kohn-Sham energy function, Ha |
| $\delta_{ij}$ | orthonormality relation, $\langle \phi_i | \phi_j \rangle$ |
| $\mu_i$ | fictitious mass, a.u. |
| $\Omega$ | volume of the unit cell, a.u.$^3$ |
| $\theta_i$ | angle of incidence, ° |
| $\theta_c$ | critical angle, ° |
| $\omega$ | frequency of the incident energy, Hz |
| $\mathcal{L}$ | Lagrangian, Ha |
| $\Psi$ | electronic wave function |
| $\mathcal{H}_e$ | Hamiltonian of the electron, Ha |
| $\phi$ | electronic orbital |
| $\lambda_{tw}$ | thermal wave length, $m$ |

# 1. Introduction

As the integrated circuits (ICs) are being miniaturized, complementary metal-oxide-semiconductor (CMOS) has step into the fabrication era of 22-nm for the central processing unit (CPU) in 2011 by Intel [1]. Transistor size and structure plays a crucial role in fulfilling the Moore's Law [2]. With the continuing improvement of manufacturing technology, the execution efficiency is also being improved. The 22-nm processors, which are known as 3-D Tri-Gate transistor uses three gates wrapped around the silicon channel in a 3-D structure, bring unprecedented computational performance, and accompany tremendous heat generation. Hence, the atomic scale thermal and thermoelectric transport should be considered with the miniaturization process. To understand the atomic scale thermal characterization and the control of materials species, interfaces, and structures are of the prime importance. In addition, the International Technology Roadmap for semiconductors (ITRS) 2006 Front End Process Update indicates that equivalent physical oxide thickness will not scale below $0.5\ nm$, which is approximately twice the diameter of a silicon atom. Consequently, to probe thermal transport would produce far reaching influence to the future development of thermal management of ICs.

In the manufacturing process, it is inevitable to leave gaps/defects inside the transistor packages at the length of the atomic scale. Thermal radiation coexists with thermal conductance that occurs at such confined geometrical structures. Due to the ubiquitous gaps in synthetic materials, especially in the fabricated ICs, understanding and predicting the heat transfer between two bodies separated by a distance of atomic scale have been a key issue in both theoretical and application points of view [3]. In addition, the thermodynamics properties of nanoscale thin films could significantly differ from that of the bulk materials [4]. Thermal energy transport is no longer a pure conceptual phenomenon but has become a crucial topic in the field of thermal management of micro-electro-mechanical systems (MEMS) and nano-electro-mechanical systems (NEMS) [5].

Because of the existence of temperature gradient in the materials, the difference of vibration energies between adjacent atoms causes them to collide with each other, which leads to diffusive transfer of kinetic energy through particles inside the materials. Heat transfer occurs spontaneously from materials at higher temperatures to that at lower temperatures. Energy carriers, such as phonons [6] and free electrons [7], play crucial rules in solid state heat conduction. As for thermal radiation, heat transfer is realized by electromagnetic radiation generated by the motion of charged particles in the matter. The spectrum emissive power of blackbody radiation is a well-understood physical phenomenon that depends on the object's temperature and obeys Plank's law. The Stefan-Boltzmann's law can be used to obtain emissive power of blackbody surface. However, it is only true for objects distances which are sufficiently large compared with the thermal wave length $\lambda_{tw}$ given by Wien's law. The typical wavelengths involved in thermal radiation are in the order of microns [8]. For the heat transfer with an interfacial distance of angstroms to nanometers, the radiation mechanism may change dramatically [9]. When two bodies are separated by a distance that is comparable or much shorter than the dominant emitted wavelength, the validity of the classical radiative transfer equations is challenged, as the wave nature of thermal radiation needs to be taken into account.

Electronic charge and thermal energy transport in nanometer scale and ultra-short time scale (from femtoseconds to nanoseconds) have drawn attentions from scholars and have been studied extensively in the past decades. For heat conduction, a unified constitutive equation covering the transport behaviors of diffusion, phonon electron interaction and pure phonon scattering was proposed by Tzou in 1995 [10]. Jia *et al.* [11] performed a numerical study of heat flux of a nanofluid system showed the thermal conductivity increases with the nanoparticle volume fraction. Different empirically predefined inter-atomic potentials were used to study the thermal dynamical properties of water by Mao [12]. For thermal radiation, a bimaterial atomic force microscope cantilever was used by Narayanaswamy *et al.* [13] to obtain the "heat transfer-distance" curves between a sphere and a substrate. Their results showed enhancement of heat transfer that exceeded that predicted by Planck's blackbody radiation theory. The surface phonon polariton was the factor that caused thermal fluctuations of the electromagnetic field. Two parallel glass surfaces were employed by Hu *et al.* [14] to measure the radiative heat flux across the micro gaps. Volokitin and Persson [15] studied heat transfer between the two parallel semi-infinite bodies separated by sub-wavelength distance via electromagnetic interaction. As it was reported in the literature, with the interfacial distance decreasing, the heat transfer increased dramatically. But the physical mechanism behind such increase still remains to be undiscovered. Kittel *et al.* [16] reported a measurement to obtain the heat transfer rate with the distance between microscope tip and sample. However, the measured results differed distinctly from the divergent behavior predicted by standard macroscopic fluctuating electrodynamics. They interpreted the results in terms of finite microscopic correlations inside the materials.

Comparing with experimental investigation, theoretical reasoning and numerical simulation can obtain results under extreme conditions that experiments cannot achieve. Molecular dynamics (MD) is a computer simulation of physical movements of atoms and molecules. The history of MD can date back to the mid-1950s when the first computer simulations on simple systems are performed [17]. In classical MD [18], with the help of the predefined empirical potential function, force field interaction between atoms, molecules, and larger clusters, can be obtained and used to calculate their motions with time. Consequently, trajectories of these molecules and atoms can be calculated via numerically solving the Newton's equations of motion. However, due to the dependence of predefined potential function to the forces acting on atoms and molecules, the accuracy and application are restricted by the available potential functions. In addition, to tackle the problems, we are facing on thermal transport issues in atomic scale via a comprehensive simulation of the thermal radiation involving electromagnetic fluctuation, for the shortcoming that classical MD fails to take dynamical electrons impact into account. The implementation of first-principles MD simulation [19] offers us an effective alternative. In the first-principles approach, forces acting on nuclei can be directly computed form surrounding electronic structure calculation via the density functional theory, and then trajectories of motion can be generated.

The first-principles based simulations of atomic scale interfacial thermal transport have not been widely investigated and documented. Luo and Lloyd [20] pioneered the first-principles non-equilibrium MD simulation of in different combinations of multi-thin-layers under thermal gradient. Later, Koker [21] extended the equilibrium MD approach to obtain the thermal conductivity based on the kinetic conductivity relation, with phonon lifetimes, group velocities, and heat capacities by combining the first-principles MD and lattice dynamics. The thin layers in Ref. [20] were placed close to each other and no distance between the neighboring layers. As aforementioned, the existence of interfacial gaps is unavoidable in reality. In this work, the first-principles MD simulation of the atomic scale energy transport via heat conduction and thermal radiation in spatial and temporal perspective is carried out. Transport phenomena occurring between silicon and/or germanium layers with gaps are investigated. Silicon is chosen because it serves as a principal component of most semiconductor devices, most importantly in the ICs and microchips. On the other hand, germanium is an important semiconductor material used in transistors, solar cells [22] and other electronic devices. In addition, the silicon-germanium alloys are rapidly becoming an important semiconductor for high-speed ICs, because of the faster transmission properties of Si-SiGe junctions than the pure silicon [23]. In this work, thin semiconductor layers with variable gap distance with an interval of lattice constant increment of the simulated materials are set up. Comparisons of temperature evolution curves under different gaps are made. Thermal transport between the same and different kinds of materials under different temperature combinations will be investigated.

## 2. Computational Methods

### 2.1 First-Principles Theory

First-principles MD of the Car-Parrinello method based on the density functional theory within local density approximation of the exchange correlation function is employed in this work. In Car-Parrinello MD, calculations are based on approach to obtain the inter-atomic force constants deriving from density functional theory. It bears with advantage of being free from the adjustable parameters and has a general applicability. A brief derivation of the simulation method from the classical MD to the first-principles MD will be described here. More detailed theoretical framework of Car-Parrinello molecular dynamics can be found in the original work [24].

Assuming a system of N particles vibrating under the influence of a potential $U$, we can describe the particles by their positions $\boldsymbol{R}$, velocities $\dot{\boldsymbol{R}}$ and masses $M$. The union of all the positions $\{\boldsymbol{R}_1, \boldsymbol{R}_2, \ldots, \boldsymbol{R}_N\}$ is denoted as $\boldsymbol{R}^N$ and the potential, which is a function of the positions only, is denoted as $U(\boldsymbol{R}^N)$. Thus, the system can be expressed in terms of Lagranian equation

$$\mathcal{L}(\boldsymbol{R}^N, \dot{\boldsymbol{R}}^N) = \sum_{I=1}^{N} \frac{1}{2} M_I \dot{R}_I^2 - U(\boldsymbol{R}^N) \qquad (1)$$

The associated Euler-Lagrange equation is

$$\frac{d}{dt}\frac{\partial \mathcal{L}}{\partial \dot{R}_I} = \frac{\partial \mathcal{L}}{\partial R_I} \qquad (2)$$

The distinct difference between classical MD and first-principles MD is the source of potential. In classical MD, the empirical potential functions are usually obtained by fitting method under prescribed conditions [25] or from experimental approaches [26]. Because of the transferability, accuracy and failure to take full account for influence of electrons in the classical MD, the simulation results may not objectively reflect and reveal fundamental chemo-physical process comparing with the first-principles MD.

When it comes to CPMD, let's focus on the Kohn-Sham method of density functional theory first [27]. The ground state energy of an interacting system of surrounding electrons with classical nuclei fixed at position $R_I$ is

$$\min_{\Psi_0} \{\langle \Psi_0 | \mathcal{H}_e | \Psi_0 \rangle\} = \min_{\{\phi_i\}} E^{KS}\{\phi_i\} \qquad (3)$$

By making simplification of replacing the minimization with respect to all possible many-body wavefunctions $\{\Psi\}$ with a minimization with a set of orthonormal one-particle functions, we can obtain the minimum of the Kohn-Sham energy as

$$\underbrace{min}_{\{\phi_i\}} E^{KS}\{\phi_i\} = T_s[\{\phi_i\}] + \int V_{ext}(r)n(r)dr + \frac{1}{2}\int V_H(r)n(r)dr + E_{xc}[n] + E_{ions}(R^N) \quad (4)$$

where $minE^{KS}$ is an explicit function of the set of Kohn-Sham orbitals $\{\phi_i\}$, which should satisfy the orthonormality relation that $\langle\phi_i|\phi_j\rangle = \delta_{ij}$. The charge density $n(r)$ is $\sum_i^{occ} f_i|\phi_i(r)|^2$.

As indicates in Eq. (1), in classical mechanics the force on nuclei is obtained from the derivative of a Lagrangian with respect to the nuclear positions. Similarly, by considering the extended Kohn-Sham energy function $\varepsilon^{KS}$ to be dependent on $\Phi_i$ and $R^N$, we can construct a Lagrangian functional derivative with respect to the orbitals, which can also be interpreted as classical fields. Therefore, in CPMD:

$$\mathcal{L}_{CP}\left[R^N, \dot{R}^N, \{\Phi_i\}, \{\dot{\Phi}_i\}\right] = \sum_I \frac{1}{2} M_I \dot{R}_I^2 + \sum_i \frac{1}{2}\mu_i\langle\dot{\Phi}_i|\dot{\Phi}_i\rangle - \varepsilon^{KS}\left[\{\Phi_i\}, R^N\right] \quad (5)$$

where the atomic mass is denoted as $M_I$, the nuclear degree of freedom is denoted as $\dot{R}_I$, and the electronic one-particle orbital is denoted as $\dot{\Phi}_i$. $\mu_i$ is the fictitious mass or inertia parameter assigned to the orbital degree of freedom.

Transforming from the quantum mechanical time-scale separation of fast electronic and slow nuclear motion into a classical mechanical adiabatic energy scale separation in the frame work of dynamical system theory, the potential (force) required in CPMD simulation is the partial derivative of the Kohn-Sham energy with respect to independently variables $\Phi_i$ and $R^N$. Therefore, the corresponding Newtonian equations of Car-Parrinello equations of motion are obtained from the associated Euler-Lagrange equations:

$$M_I \ddot{R}_I(t) = -\frac{\partial \varepsilon^{KS}}{\partial R_I} = -\frac{\partial[E^{KS} + \sum_{ij}(\langle\Phi_i|\Phi_j\rangle - \delta_{ij})]}{\partial R_I} = -\frac{\partial E^{KS}}{\partial R_I} + \sum_{ij}\Lambda_{ij}\frac{\partial\langle\Phi_i|\Phi_j\rangle}{\partial R_I} \quad (6)$$

$$\mu_i \ddot{\Phi}_i(t) = -\frac{\delta \varepsilon^{KS}}{\delta \langle \Phi_i|} = -\frac{\delta[E^{KS} + \Sigma_{ij}(\langle \Phi_i|\Phi_j\rangle - \delta_{ij})]}{\delta \langle \Phi_i|} = -\frac{\delta E^{KS}}{\delta \langle \Phi_i|} + \Sigma_j \Lambda_{ij}|\Phi_j\rangle \tag{7}$$

In CPMD code, pseudopotential plane wave method is employed in the simulation algorithm. In a modeled system, a periodically repeated unit cell is defined by the lattice vectors $\boldsymbol{a}_1$, $\boldsymbol{a}_2$, and $\boldsymbol{a}_3$. Combining with the scaled coordinates $s$, we can express the established system as

$$\boldsymbol{r} = s\boldsymbol{h} = s[\boldsymbol{a}_1 \quad \boldsymbol{a}_2 \quad \boldsymbol{a}_3] \tag{8}$$

Reciprocal lattice vectors $\boldsymbol{b}_i$ are defined as

$$\boldsymbol{b}_i \cdot \boldsymbol{a}_j = 2\pi \delta_{ij} \tag{9}$$

Periodic boundary condition are established by using

$$\boldsymbol{r}_{pbc} = \boldsymbol{r} - \boldsymbol{h}[h^{-1}r]_{NINT} \tag{10}$$

where $[h^{-1}r]_{NINT}$ is the nearest integer value. With the assumed periodicity, an orthonormal basis is built by plane waves as

$$f_G^{PW}(\boldsymbol{r}) = \frac{1}{\sqrt{\Omega}} e^{i\boldsymbol{G} \cdot \boldsymbol{r}} = \frac{1}{\sqrt{\Omega}} e^{2\pi i \boldsymbol{g} \cdot \boldsymbol{s}} \tag{11}$$

where $\Omega$ is the volume of the unit cell. Plane waves represent the periodic part of the orbitals in form of basis set, which realizes the purpose that the periodic lattice produces a periodic potential and then imposes the same periodicity on the density.

The employment of norm conserving pseudopotential approach provides an effective and feasible way to performing calculations on the complex systems (liquid and solid state) using plane wave basis sets. By taking the chemically active valence electrons and eliminated inert core electrons, the pseudopotentials are constructed. The properties of additivity and transferability should be satisfied for the pseudopotentials. Thus, the purpose of obtaining smooth pseudo orbitals to describe the valence electrons and keeping the core and valences states orthogonal in an all electron framework is realized.

## 2.2 Modeling and Simulation Details

With the rapid development of cutting edge thin film growth technology, it is possible to arrange different materials at the atomic level and to fabricate the thin-film structures with strong size-effect from angstroms to hundreds nanometers. In this work, the numbers of atoms with 4(length) × 4(width) are built into square layers of 2 (thickness) atoms aligned in the third direction for each layer. Periodic boundary conditions are applied in all three dimensions. In the case of silicon-silicon-layer combination, the lattice constant of 5.431 Å is used in Cartesian coordinate modeling setting up. In the case of silicon-germanium-layer combination, the change is lattice constants of 5.431 Å for silicon and 5.658 Å for germanium, respectively. Both the silicon and germanium are set as diamond cubic crystal structure, which is a repeating pattern of 8 atoms. To take a general integration of adjacent layers (4(length) × 4(width) × (2 + 2)(thickness)) and the gap discrete layers (4(length) × 4(width) × (2 + 2 + $x$)(thickness), where $x$ refers to the ratio of gap distance to lattice constant), the supercell symmetry is set as orthorhombic. Figure 1 (rendered using VMD [28]) shows the silicon-germanium-layer combination with an interfacial distance equal to the germanium lattice constant.

Our simulations are based on the latest CPMD v.3.15.3 [29]. Each MD simulation case starts from the wavefunction optimization to perform electron structure calculation of the ensemble to make sure that the consequent CPMD simulation physically run near the Born-Oppenheimer surface. Simulation time step is set to be $4.0\ a.u.$ $(1\ a.u. = 0.024,188,842,8\ fs)$. At the first temperature control stage, canonical or NVT ensemble is employed to ensure the ensemble has a well-defined temperature. Nose-Hoover thermostats [30] are imposed on ions and electrons for each degree of freedom. The energy of endothermic and exothermic processes is exchanged with the clapped thermostats. Characteristic frequency of electrons and ions are chose to be 10,000 and 2,500, respectively. After the temperature of the entire system reaching to steady state, thermostats are removed to keep the system as a microcanonical or NVE ensemble and an adiabatic stage starts to take place with no heat exchange between the system and its boundary. The only process is the exchange of potential and kinetic energy inside the ensemble, with total energy being conserved. The norm-conserving pseudopotentials are used for minimizing the size of the plane wave basis to generate the same charge density as the real full electrons. For the Trouiller-Martins [31] and Stumf-Gonze-Scheffler [32] pseudopotentials are used for silicon and germanium layers, respectively. The thermal energy exchange between the neighboring layers and potential energy of atoms in the layers will be considered. In the cases of interfacial distance varying with an interval of lattice constant increment of simulated materials, because the gaps between layers can be regarded as vacuum, and the thermal energy exchange is realized from heat conduction to thermal radiation.

## 3. Results and Discussions

### 3.1 Pure Material Combinations

#### 3.1.1 Silicon-Silicon-Layer Combinations

The silicon-silicon-layer at 400 K versus 100 K under different interfacial distance (closely contact, 1.0 / 2.0 times lattice constant of silicon) is modeled. Since the neighboring layers belong to the same material and the same kind of structure, the equilibration time from the removal point of the thermostats to the thermal equilibrium state is very short although such equilibration time duration renders a tendency of increasing with the increasing interfacial distance (Table 1).

Simulation results show equilibration time of the closely contacted layers is the shortest, as seen in Fig. 2. The thermal transport phenomenon arising from these well matched layers attributes to heat conduction. Heat conduction within non-metallic solids occurs primarily by the propagation of elastic waves associated with the displacement of atoms from their lattice sites. Phonons are defined from these the spatially localized, quantized unit of propagating vibration waves. In semiconductors, like silicon, phonons are also the major energy carriers. Because the free electron density in a normal semiconductor is much lower than that in a metal, the participation of electrons can be neglected. For the layers established in our simulation, the length scale of the layers is approximate to lattice constant of silicon (2 atoms in the thickness direction), which is far less than the phonon mean free path $(100 - 1000 \text{ Å})$. Thus, the thermal transport mechanism is ballistic conduction. The vibration energy of phonons transfers ballistically from one layer to another. In other words, the phonons keep along straight-line trajectories without destruction or redirection. Simulation result show the equilibration time is about 0.2800 ps (Fig. 2). Since the distance between the geometric centers of the neighboring silicon layers is set to be $5.431 \text{Å}$ and the speed of sound of silicon is $5860 - 8480$ m/s [33], the estimated time it takes for one way transport should be in the range of $0.0640 - 0.0927$ ps. Taking the back and forth thermal communication between the contacting layers before the final thermal equilibrium state, the longer equilibration time of 0.2800 ps is reasonable in our simulation result.

When the two layers are separated and set with different interfacial distance, simulation results in Figs. 3 and 4 show that the equilibration time for the layers from the initial non-equilibrium to thermal equilibrium states gradually becomes longer as the interfacial distance increases (Table 1). The essential distinction between these latter cases (Figs. 3 and 4) and the first case (Fig. 2) is that the thermal energy transport transform from the heat conduction to the thermal radiation with the enlargement of the layers interfacial gap. The dipole oscillation of the charges leads to the production of electromagnetic radiation. In the generated coupling electric and magnetic fields, radiating thermal energy emits from the body through its surface boundary. There are two modes for the electromagnetic radiation: far-field propagation electromagnetic modes and near-field evanescent mode that decays quickly over submicron length scales. In our simulated cases, the interfacial distance is at nanoscale, which is far less than the decay length of the near-field modes. Evanescent waves and phonon tunneling are responsible for the near-field energy transfer. The evanescent wave field of higher temperature layer excites the charges within lower temperature layer and dissipates its energy from high-temperature layer to low-temperature layer. In addition, the surface waves are hybrid modes that generate from coupling of electromagnetic field and mechanical oscillation of energy carriers inside the materials. For silicon whose energy carriers mainly composed with phonons, the hybrid mode is named surface phonon-polariton. Due to the spatial difficulties of propagating waves caused by the enlarging of interfacial distance, equilibration time becomes correspondingly longer due to weakened energy communication occurring on the surface of each layer. The simulation results are consistent with the theoretical analysis.

### 3.1.2 Germanium-Germanium-Layer Combinations

For the models of germanium-germanium-layer combinations, simulations are carried out under almost the same conditions as Section 3.1.1. Energy transport processes between two layers of germanium atoms at 400 K and 100 K under different interfacial distances are studied. The simulation results (Table 2) show longer equilibration time for both heat conduction and thermal radiation in the cases of germanium than that of the silicon. For the heat conduction (Fig. 5), the longer equilibration time for germanium agrees with the macroscale property that germanium has a relatively lower thermal conductivity of $60.2\ W \cdot m^{-1} \cdot K^{-1}$ than that of $149\ W \cdot m^{-1} \cdot K^{-1}$ for silicon. As for the reason why equilibration time for thermal radiation (Figs. 6 and 7) is also longer comparing with the cases in Section 3.1.1, we will have an explanation in detail in Section 3.2.3.

### 3.2 Hybrid Materials Combinations

Each of silicon and germanium has four valence electrons. However, germanium will have more free electrons and more active thermoelectric characters than that of silicon under the same temperature, because germanium has lower band gap than that of silicon. On the other hand, the stability of silicon becomes one of the reasons that it is widely used in semiconductors at higher temperatures than germanium. To investigate the temperature and material dependence of both silicon and germanium, we carried out simulation of silicon-germanium-layer combinations at silicon (400 K) versus germanium (100 K) and silicon (100 K) versus germanium (400 K), respectively.

**3.2.1 Silicon (400 K) versus Germanium (100 K)**

Since the temperature of silicon is set higher than that of germanium, the overall temperature evolution tendencies would be to cool down the silicon layers and to heat up the germanium. Figures 8-10 show the evolution of the temperatures of the silicon and germanium layers with different separation distances. Similar to the case of silicon-silicon-layer heat conduction (Fig. 2), the silicon (400 K) - germanium (100 K) - layer combination (Fig. 8) shows steep decrease and increase of temperatures during the process from non-equilibrium to dynamic thermal equilibrium. However, in the consequent conduction process, temperatures of both the silicon and germanium layers rend very slow variation tendencies. It may due to relatively small temperature difference between the two layers. The total equilibration time of thermal transfer is about 3.096 ps (32,000 time steps). But comparing the heat conduction process of the pure silicon-silicon-layer combination (Fig. 2) with the silicon (400 K) - germanium (100 K) - layer combination (Fig. 8), one can conclude that the latter is distinctly longer. Probing from the atomic vibration point of view, the acoustic mismatch may be the main factor that leads to more difficult thermal communication for silicon-germanium-layer. In addition, mass difference is another factor that could contribute to the difference.

When it comes to radiative heat transfer, simulation results (Figs. 9 and 10) show that the slopes of for both the silicon and germanium temperatures become relatively gentle compared with the heat conduction case (Fig. 8). Additionally, the equilibration time of thermal transfer for silicon (400 K) - germanium (100 K) - layer combination tends obviously much longer than that of the silicon-silicon-layer/germanium-germanium-layer combinations. Considering thermal radiation emission from the electrodynamics point of view, the propagating and evanescent waves are emitted via the out-of-phase oscillations of charges of opposite signs. The couples of charges of opposite signs are named dipoles. The random fluctuation of charges in turn generates a fluctuating electromagnetic field, named thermal radiation field. Because of the different kinds of thermal radiation field generated from the silicon and germanium layers, thermal communication difficulties in the inter-radiation field becomes dominant factors that lead to longer equilibration time in the silicon (400 K) - germanium (100 K) - layer combination than the pure material combinations. The larger interfacial distance is set up, the longer equilibration time need for neighboring layers to reach thermal equilibrium. This is also consistent with the mechanism for results with gradually enlarged gaps in the silicon-silicon-layer combination presented in Section 3.1.

**3.2.2 Silicon (100 K) versus Germanium (400 K)**

We exchange the temperature configurations of the silicon layer and germanium layer to probe the impact of atoms' temperature difference during the thermal transport process. Like simulations performed before, in the silicon (400 K) - germanium (100 K) - layer combination model, we start our work from closely contacting layers.

The closely contacting layers thermal conduction case (Fig. 11) and the gap enlarged layers thermal radiation cases (Figs. 12 and 13) render much longer equilibration time than those at the same conditions of silicon (400 K) - germanium (100 K) - layer cases (Figs. 8-10). It shows that the thermophysical properties of nanocomposite materials depend on not only the properties of their individual constituents but also their morphology and interfacial characteristics. The heat conduction results (See Figs. 8 and 11) indicat that the thermal contact conductance in the interface is significantly affected by how the temperatures differences are imposed. As shown in our simulation results for heat conduction (Figs. 8 and 11), it is easier for the thermal energy transfer from the silicon side to the germanium side under the condition of relatively lower temperature for germanium layer. Thus, the ballistic thermal conductance becomes lower when the temperature for germanium is higher than that of the silicon. In addition, the heterogeneous effect is another factor deserving our attention. For the case of silicon (100 K) - germanium (400 K) - layer combination, the acoustic mismatch accounting for long-wavelength phonons is the main impedance in the interfacial thermal conduction. From the phonon density of states point of view, the diffusion of scattering phonons mismatch is another factor that weakens the thermal transport for silicon and germanium combinations than that of the pure material combinations.

The thermal radiation results for silicon (100 K) - germanium (400 K) - layer combination demonstrated a phenomenon: when the interfacial distance between silicon and germanium layers increases, the corresponding equilibration time to reach thermal equilibrium becomes much longer than that of the same cases for the silicon (400 K) - germanium (100 K) - layer combinations (See Tables 3 and 4 for details). As explained before, different kinds of evanescent waves are generated in the surfaces of silicon and germanium, respectively. Because of the interfacial distances are in the sub-nanometer and nanometer in our simulated cases, which are far less than the decay length of the near field modes, the vacuum gap between neighboring layers can be treated as space filled with evanescent waves along the boundaries of both facing layers.

### 3.2.3 Further Discussion on Equilibrium Time

In our simulated models, the characteristic structural length is much shorter than the photon wavelength. Thus, the wave properties of the radiation energy should be considered. Because of the quantum and vacuum fluctuations of atoms and electrons, there will be a fluctuating electrodynamical field inside the layers. In addition, according to the boundary conditions of macroscopic electrodynamics, a continuous electric and magnetic field has to exist in the spatial gap. That is to say, there is also a fluctuating electromagnetic field outside the layers produced by fluctuating internal source. The difference of equilibration time can be explained by the total internal reflection. The index of refraction of germanium is greater than that of silicon, and both of them are greater than 1 in vacuum [34]. Analyzing from the wave emission point of view for the higher temperature layer of germanium, let us describe the wave incident from germanium to vacuum as the electric field $E_{Ge \to vacuum} e^{i(k_{Ge} r - \omega t)}$, where $k_{Ge}$ denotes as the wave vector from germanium to vacuum. To simplify the three-dimension thermal radiation filed, let us assume that the waves are travelling the $x - y$ plane only. As seen in Fig. 14. The amplitude of the wave vector $k_{Ge \to vacuum} = |\mathbf{k}_{Ge \to vacuum}|$ equals to $n_{Ge}k_0$, where $k_0$ is the magnitude of the wave vector in vacuum.

Thus, the x-component of the wave vector in the surface of germanium layer is:

$$k_{x,Ge} = \sqrt{(n_{Ge}k_0)^2 - k_{y,Ge}^2} \tag{12}$$

The y-component of the wave vector can be expressed in term of the angle of incidence:

$$k_{y,Ge} = n_{Ge}k_0 \cdot sin\theta_{i,Ge} \tag{13}$$

where $\theta_i$ is the angle of incidence. Since the index of refraction of germanium is greater than 1, there will be an angle of incidence that makes the entire energy irradiating into the vacuum gap be reflected back to the germanium layer. Thus, a critical angle ($\theta_c$) at which the angle of incidence is equal to the angle for total internal reflection can be defined. At the moment the y-component of the wave vector is:

$$k_{y,Ge} = 1 \tag{14}$$

We can derivate $\theta_{c,Ge}$ as

$$\theta_{c,Ge} = sin^{-1}\frac{1}{n_{Ge}} \tag{15}$$

In our simulated system, periodic boundary condition is employed in three dimensions. Hence, the y-component of the vector is conserved because the layers are infinite along the y-direction. Similarly, the x-component of the wave vector generated from silicon to germanium can be expressed as

$$k_{x,Si} = \sqrt{(n_{Si}k_0)^2 - k_{y,Si}^2} \tag{16}$$

and the critical angle of incidence is

$$\theta_{c,Si} = sin^{-1}\frac{1}{n_{Si}} \tag{17}$$

Thus, we can conclude that $\theta_{c,Si} > \theta_{c,Ge}$ (Fig. 15). The incident waves, sourcing from germanium and silicon layers, have the same probabilities involved in all direction ($0° - 180°$). The percentage of incident waves which can be effectively transformed as evanescent waves on the surface of germanium layer will be less than that on the silicon layer, due to the relatively greater refraction number of germanium.

Reversely, peeking from the scope of energy absorption, its direction is from the vacuum to silicon/germanium layers. Here, the angle of incidence ranges from $0°$ to $90°$, which means that various incident waves can directly penetrate into the silicon/germanium layers. Furthermore, since there are more electrons circling around the germanium nuclei than that of silicon nuclei, even if the same amount of thermal energy penetrate into the two materials, it would be much easier for the germanium layer to absorb the external energy and convert it into internal energy. In other words, the temperature increasing rate for germanium (100 K) layer would be faster than silicon (100 K) layer.

Therefore, from both the energy emission and absorption point of view, owing to dual action of evanescent wave emission and absorption capability, the heat transfer rate for the silicon (400 K) - germanium (100 K) - layer combinations is greater than silicon (100 K) - germanium (400 K) - layer combinations in our simulation results, which is an agreement between the simulation results with the theoretical analysis.

As for the reason why the equilibration time for the gap of 11.316 Å is longer than that of the gap of 5.658 Å cases in both the silicon (400 K) - germanium (100 K) - layer combination and the silicon (100 K) - germanium (400 K) - layer combination, we can interpret it with the evanescent-wave coupling [35]. As aforementioned, the evanescent waves are generated on both the silicon and germanium layer surfaces. The sort of electromagnetic waves decay exponentially in electromagnetic field. The coupling phenomenon occurs by placing two electromagnetic fields close together so that the evanescent field generated by one layer does not decay much before it reaches to the neighboring layer.

Thus, for larger interfacial distance set up in the simulated model, the less coupling waves are involved in the vacuum gap. Consequently, the thermal energy transport efficiency will be lower for larger interfacial gap.

## 4. Conclusion

We performed atomic scale simulation of thermal conduction and radiation in combinations of silicon and germanium thin layers to investigate the energy transport process from non-equilibrium to equilibrium. Silicon-silicon-layer ( 400 K versus 100 K ), germanium-germanium-layer ( 400 K versus 100 K ), and silicon-germanium-layer ( 400 K versus 100 K and 100 K versus 400 K ) combinations are simulated under gradually increasing interfacial distance between the neighboring layers. The impacts of interfacial distance and material species are rendered via plotting of the temperature evolutions for the neighboring layers. From the heat conduction to thermal radiation process, equilibration time becomes longer, except for the combination of silicon (100 K) - germanium (400 K) - layer in the heat conduction case. With increasing interfacial distance, the equilibration time becomes longer. Under the condition of the pure material combination (silicon–silicon-layer and germanium-germanium-layer), the thermal communication develops significantly faster than that of the hybrid materials combinations (silicon-germanium-layer). The acoustic mismatch is the main factor that leads to difficult thermal communication for silicon-germanium-layer. In addition, mass difference and configurations of electrons surrounding nucleus are other factors. The distinct differences between silicon-germanium-layer under conditions of 400 K versus 100 K and 100 K versus 400 K indicate the interfacial thermal physical properties of nano-composite materials depend not only on the properties of their individual constituents but also their morphology and connection characteristics. Due to the different index of incidence of silicon and germanium, the evanescent waves generated by vibrations of charge dipoles are impeded under different critical angles and directly lead to longer equilibration time for high temperature germanium versus low temperature silicon combination. With the advantage of high accuracy of electron structure calculation, we believe that the first-principles simulation based on the Car-Perrinello molecular dynamics theory will serve as a novel approach to solve atomic scale thermal transport problems.

## Acknowledgment

Support for this work by the U.S. National Science Foundation under grant number CBET- 1066917 is gratefully acknowledged.

# Table caption

Table 1  Equilibration time for silicon-silicon-layer combinations under different interfacial distance

Table 2  Equilibration time for germanium-germanium-layer combinations under different interfacial distance

Table 3  Equilibration time for silicon (400 K) - germanium (100 K) - layer combinations under different interfacial distance

Table 4  Equilibration time for silicon (100 K) - germanium (400 K) - layer combinations under different interfacial distance

**Table: 1**

| Interfacial Distance (Å) | Time steps | Time (ps) |
|---|---|---|
| 0.000 | 2,900 | 0.280 |
| 5.431 | 5,100 | 0.494 |
| 10.862 | 6,800 | 0.658 |

**Table: 2**

| Interfacial Distance (Å) | Time steps | Time (ps) |
|---|---|---|
| 0.000 | 6,500 | 0.629 |
| 5.658 | 18,000 | 1.742 |
| 11.316 | 20,000 | 1.936 |

**Table: 3**

| Interfacial Distance (Å) | Time steps | Time (ps) |
|---|---|---|
| 0.000 | 32,000 | 3.096 |
| 5.658 | 34,000 | 3.289 |
| 11.316 | 36,000 | 3.483 |

**Table: 4**

| Interfacial Distance (Å) | Time steps | Time (ps) |
|:---:|:---:|:---:|
| 0.000 | 194,000 | 18.770 |
| 5.658 | 57,000 | 5.515 |
| 11.316 | 74,000 | 7.160 |

# Figure Caption:

Fig. 1  Simulation box of silicon-germanium-layer combination with the Interfacial distance equals the lattice constant of germanium

Fig. 2  Silicon-silicon-layer combination (attached to each other)

Fig. 3  Silicon-silicon-layer combination (interfacial distance: 1.0 lattice constant of silicon)

Fig. 4  Silicon-silicon-layer combination (interfacial distance: 2.0 lattice constant of silicon)

Fig. 5  Germanium-germanium-layer combination (attached to each other)

Fig. 6  Germanium-germanium-layer combination (interfacial distance: 1.0 lattice constant of germanium)

Fig. 7  Germanium-germanium-layer combination (interfacial distance: 2.0 lattice constant of germanium)

Fig. 8  Silicon-germanium-layer combination (400 K vs. 100 K, attached to each other)

Fig. 9  Silicon-germanium-layer combination (400 K vs. 100 K, interfacial distance: 1.0 lattice constant of germanium)

Fig. 10 Silicon-germanium-layer combination (400 K vs. 100 K, interfacial distance: 2.0 lattice constant of germanium)

Fig. 11 Silicon-germanium-layer combination (100 K vs. 400 K, attached to each other)

Fig. 12 Silicon-germanium-layer combination (100 K vs. 400 K, interfacial distance: 1.0 lattice constant of germanium)

Fig. 13 Silicon-germanium-layer combination (100 K vs. 100 K, interfacial distance: 2.0 lattice constant of germanium)

Fig. 14 Periodic display of the silicon-germanium-layer combination (the interfacial distance equals the lattice constant of germanium)

Fig. 15 Schematic shown the incident waves, emitted waves, evanescent waves during the electromagnetic waves emission process

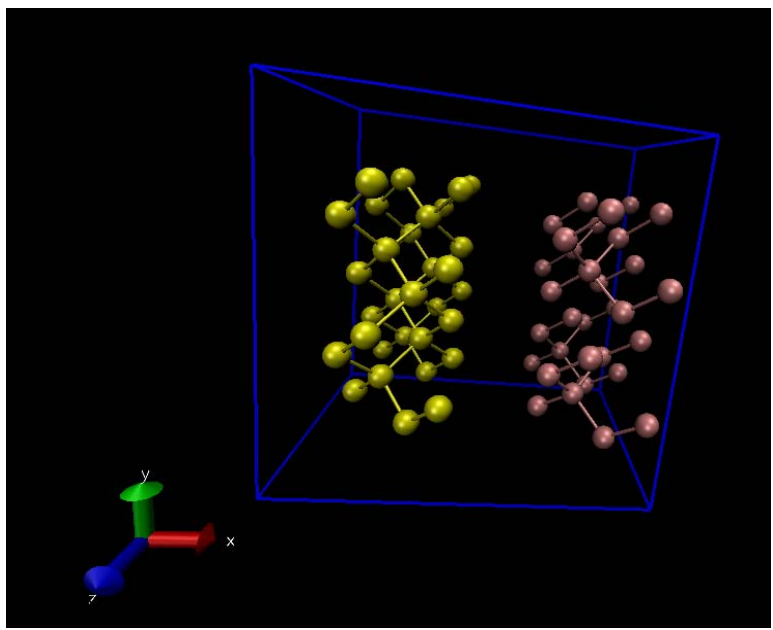

**Fig. 1**

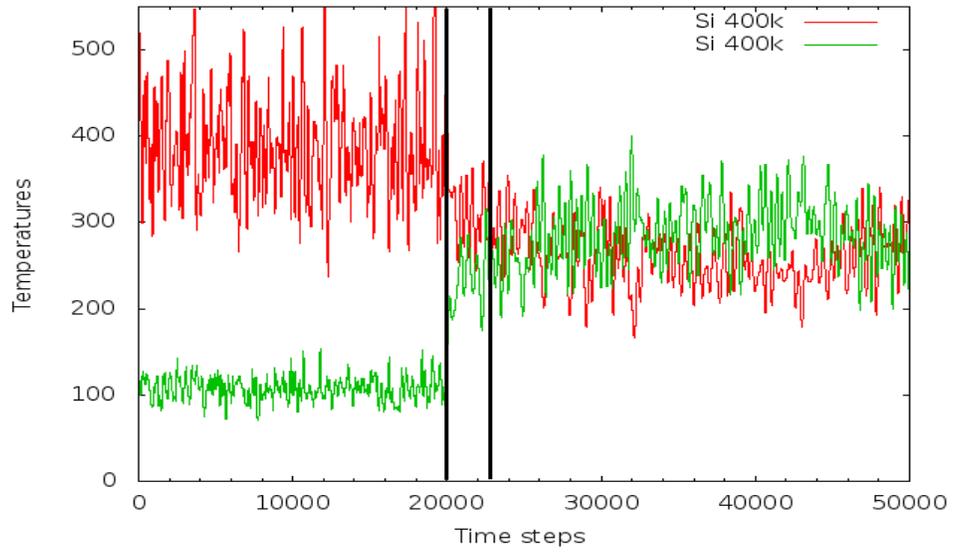

**Fig. 2**

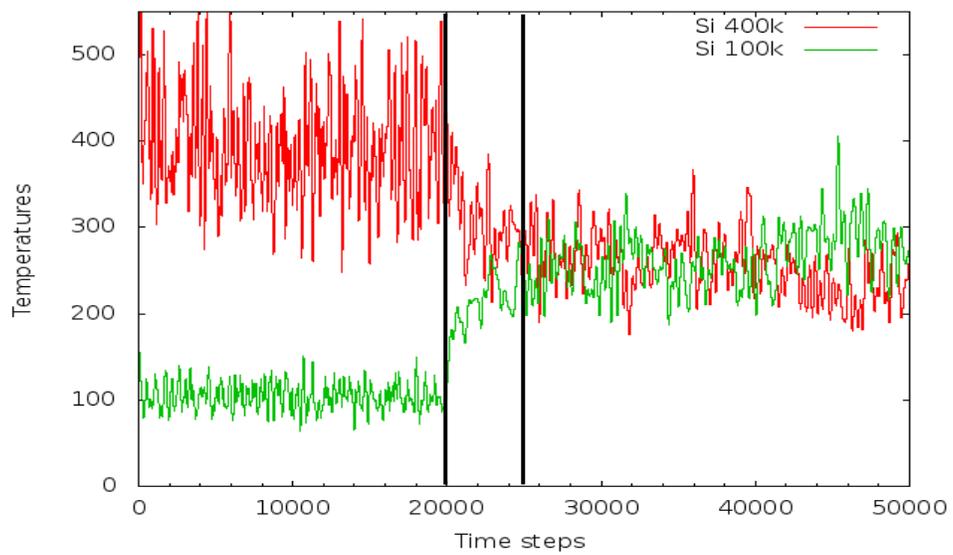

**Fig. 3**

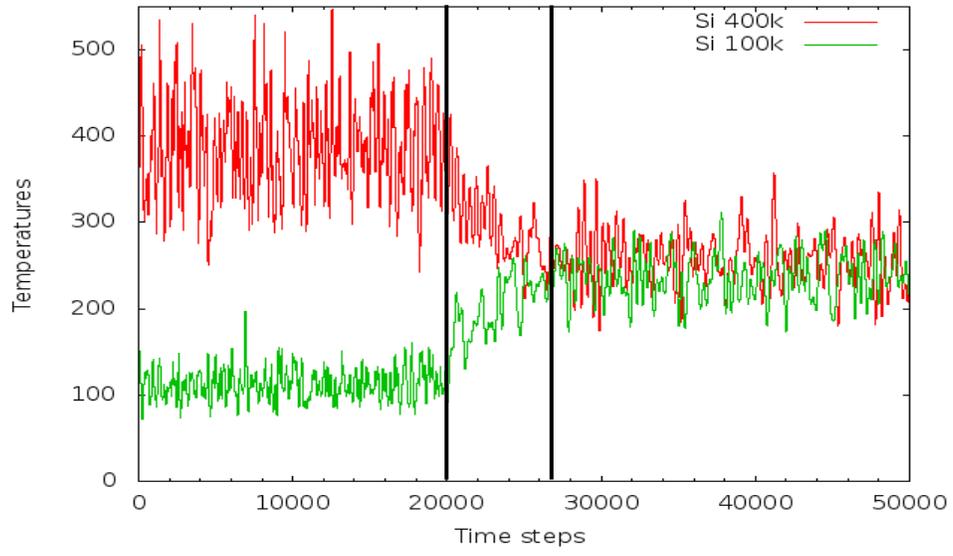

**Fig. 4**

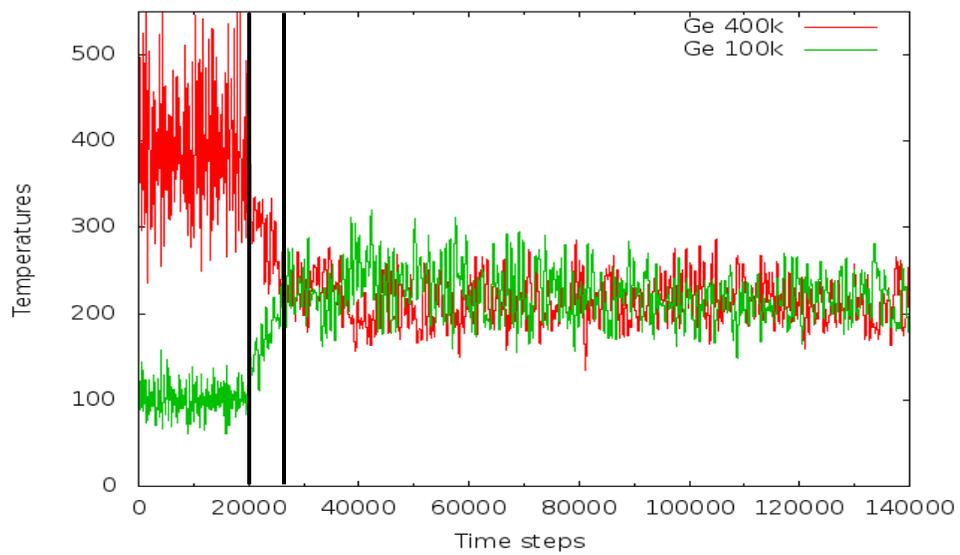

**Fig. 5**

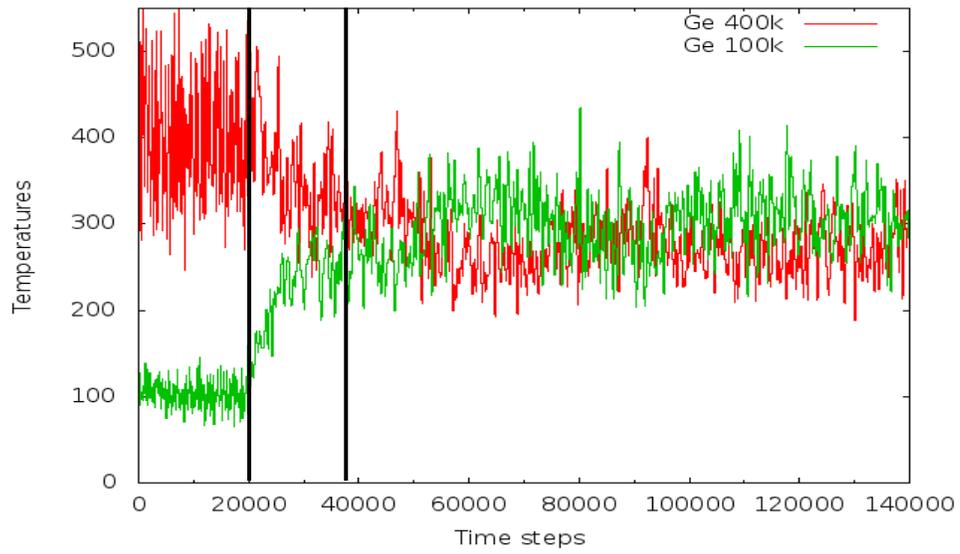

**Fig. 6**

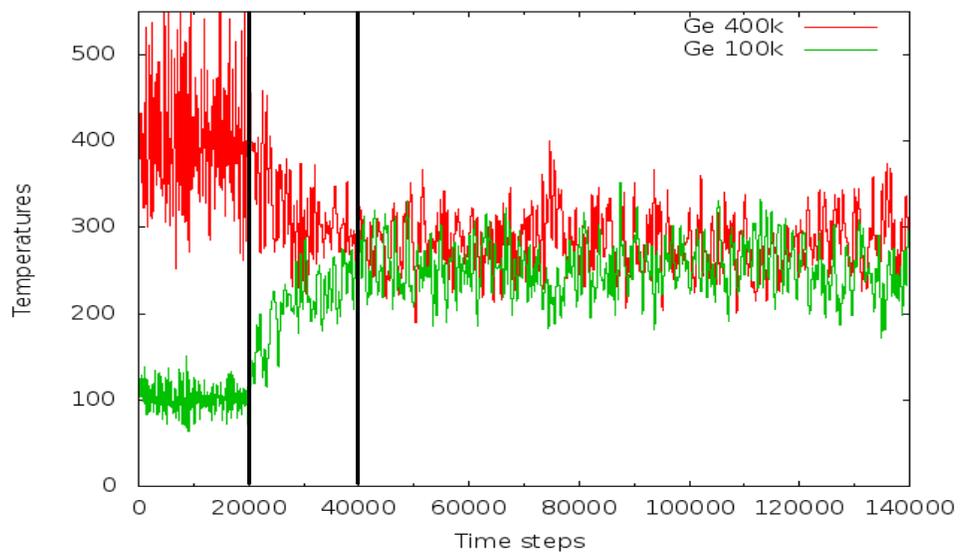

**Fig. 7**

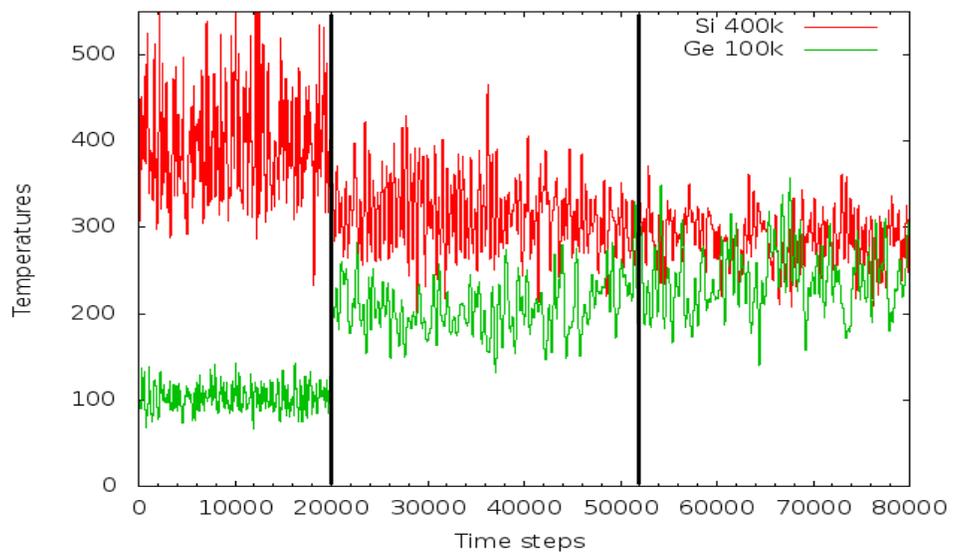

**Fig. 8**

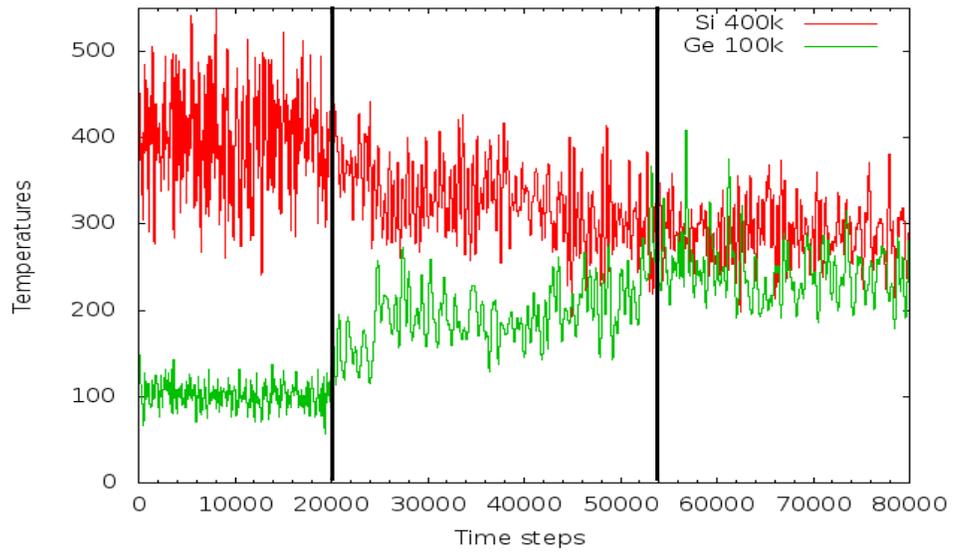

**Fig. 9**

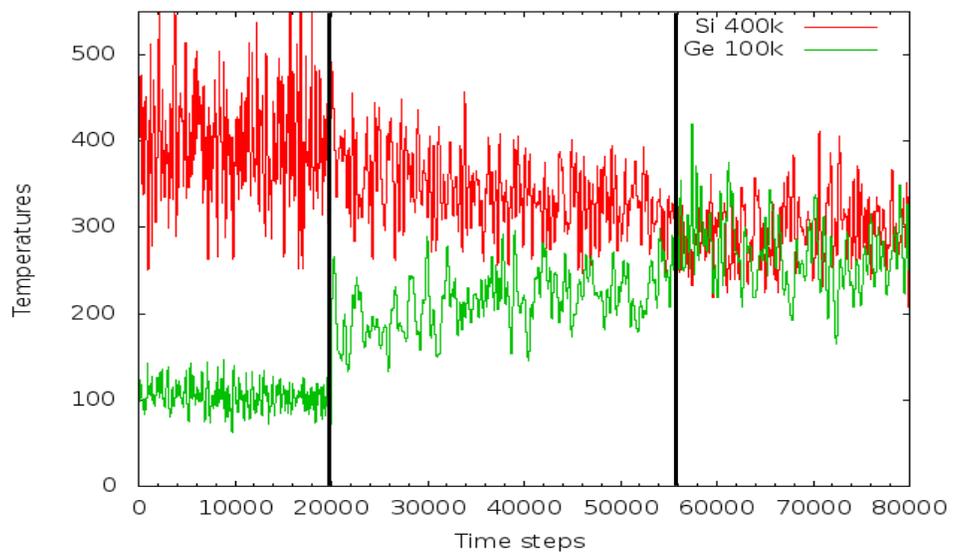

**Fig. 10**

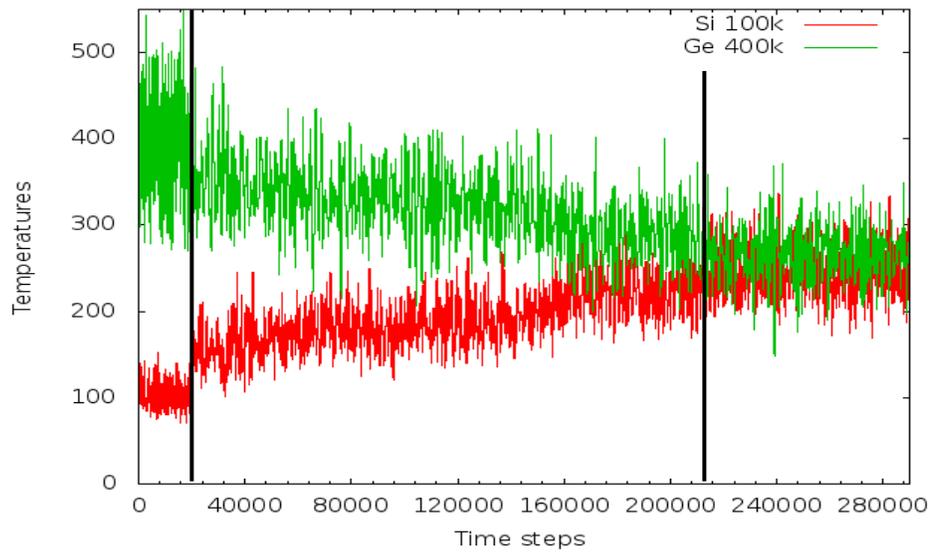

**Fig. 11**

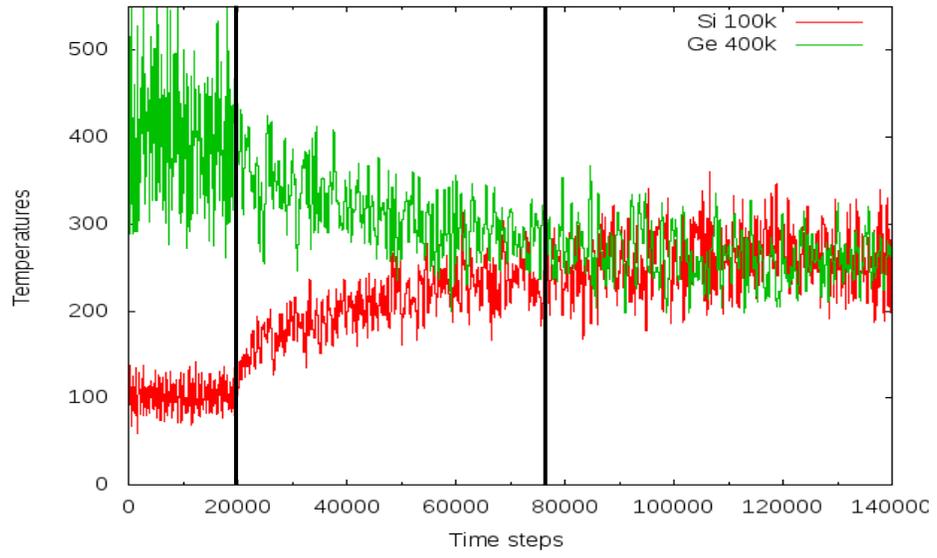

**Fig. 12**

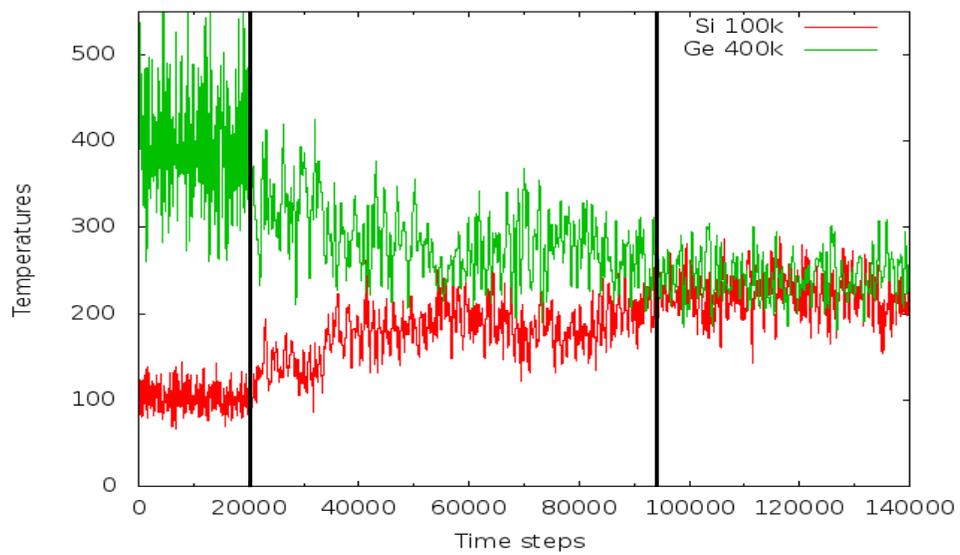

**Fig. 13**

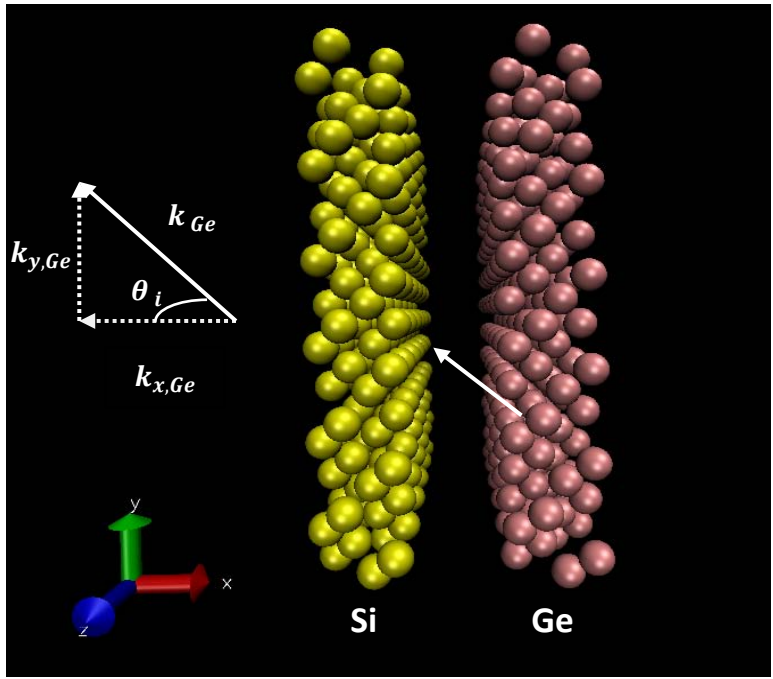

**Fig. 14**

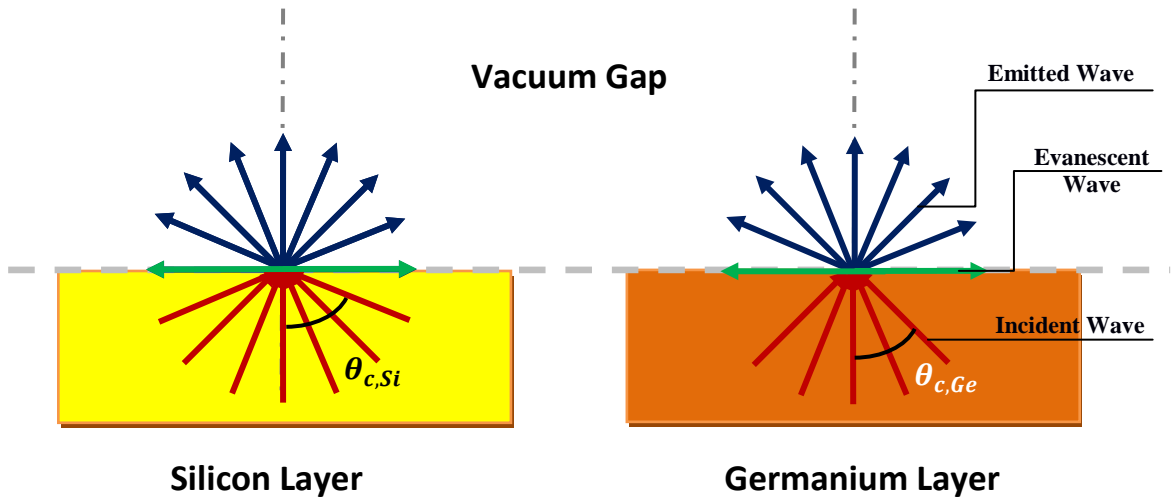

**Fig. 15**